# Community-Informed AI Models for Police Accountability[1]

Version: April 22, 2026

Benjamin A.T. Graham[a,*], José J. Alcocer[n], Lauren Brown[b], Georgios Chochlakis[c,d], Morteza Dehghani[b,e,l], Raquel Delerme[f], Brittany Friedman[f], Ellie Graeden[g], Preni Golazizian[d], Rajat Hebbar[c,i], Parsa Hejabi[d], Aditya Kommineni[c,i], Harry G. Muttram[m], Mayagüez Salinas[j], Michael Sierra-Arévalo[k], Jackson Trager[l], Nicholas Weller[m], and Shrikanth Narayanan,[d,i]

**Author Affiliations**

[a]Department of Political Science and International Relations, University of Southern California, Los Angeles, USA 90007
[b]School of Public Policy, University of Southern California, Los Angeles, USA 90007
[c]Signal Analysis and Interpretation Laboratory (SAIL), University of Southern California, Los Angeles, USA 90007
[d]Department of Computer Science, University of Southern California, Los Angeles, USA 90007
[e]Brain and Creativity Institute, University of Southern California, Los Angeles, USA 90007
[f]Department of Sociology, University of Southern California, Los Angeles, USA 90007
[g]Center for Global Health Science and Security, Georgetown University, Washington, DC USA 20057
[i]Department of Electrical and Computer Engineering, University of Southern California, Los Angeles, USA 90007
[j]The Lewis Registry, Los Angeles, CA 90049.
[k]Department of Sociology, The University of Texas at Austin, Austin, USA 78712
[l]Department of Psychology, University of Southern California, Los Angeles, USA 90007
[m]Department of Political Science, University of California Riverside, Riverside, USA 92521
[n]Harvard Law School, Harvard University, Cambridge, USA, 02138
[*]To whom correspondence should be addressed: Email: benjamag@usc.edu

[1] A prior (2024) version of this paper was titled "A Multi-Perspective Machine Learning Approach to Evaluate Police-Driver Interaction in Los Angeles"

**Abstract**[2]
Face-to-face interactions between police officers and the public affect both individual well-being and democratic legitimacy. Many government-public interactions are captured on video, including interactions between police officers and drivers captured on bodyworn cameras (BWCs). New advances in AI technology enable these interactions to be analyzed at scale, opening promising avenues for improving government transparency and accountability. However, for AI to serve democratic governance effectively, models must be designed to include the preferences and perspectives of the governed. This article proposes a community-informed, approach to developing multi-perspective AI tools for government accountability. We illustrate our approach by describing the research project through which the approach was inductively developed: an effort to build AI tools to analyze BWC footage of traffic stops conducted by the Los Angeles Police Department. We focus on the role of social scientists as members of multidisciplinary teams responsible for integrating the perspectives of diverse stakeholders into the development of AI tools in the domain of police – and government – accountability.

---

[2] **Author Contributions:** Graham led the writing and coordinated the project; Alcocer conducted the empirical analysis of the survey responses with primary guidance from Weller; Brown, Delerme, Friedman, Graham, and Sierra-Arévalo conducted interview and focus group research on community perspectives and Delerme observed officer training; Narayanan led development of our multimodal modeling approach with contributions from Chochlakis, Hebbar, and Kommineni; Dehghani led development of our multiperspective modeling approach with contributions from Golazizian, Hejabi, and Trager; Hejabi developed the multiperspective video annotation platform; Graeden led our data integration and data security strategy; Muttram conducted the empirical analysis of the annotation data with primary guidance from Weller; Salinas established stakeholder relationships and shaped overall project design; Trager led development of our annotation manual and annotator training; Weller led our survey research. All authors contributed ideas throughout the development of the project and prose throughout the paper.

## Introduction

Body-worn cameras (BWCs) were introduced into American policing to increase transparency and accountability and promote the safety of both officers and the public. Unfortunately, there is little evidence to date that those goals have been achieved (Lum et al. 2020; National Institute of Justice 2022), in part because of the overwhelming cost and logistical challenges of manually reviewing vast amounts of footage (Camp and Voigt 2025; Newell 2021).

Recent technological advances in artificial intelligence (AI) create new possibilities for automated analysis of BWC recordings and other video data (Camp et al. 2024; Camp and Voigt 2025; Rosas-Smith et al. 2025; Voigt et al. 2017; Srbinovska et al. 2025; White, Watts, and Malm 2026). Large-scale automated analysis can enable both the kind of systematic oversight originally envisioned and also rapid learning about the impacts of policies and tactics. Developing AI tools for the purposes of improving government performance and facilitating the accountability of police officers and other government officials responsible to the communities they serve requires not only advances in engineering and computer science, but also social-scientific innovations with respect to the design of AI tools that reflect the perspectives and priorities of the governed.

Our goal in this paper is to establish general principles for the development of AI tools in the context of government transparency and accountability, with a particular focus on law enforcement. These principles should guide both academic and industry practitioners who develop such tools and police departments and government agencies who must evaluate and decide whether to adopt these new technologies. In doing so, we discuss an ongoing project analyzing policing in Los Angeles.

Democratic accountability is important in designing AI tools to facilitate any aspect of government oversight but is particularly important in the context of American policing. Police departments frequently represent the largest line item in local government budgets and policing is a domain in which the inequitable application of government power is well-documented (Braga, Brunson, and Drakulich 2019; Soss and Weaver 2017). Almost half of Americans – and a majority of Americans of color – believe that the police are not "generally held accountable for misconduct" when misconduct occurs (Ekins 2016). Past efforts to bring algorithmic and AI tools into law enforcement, corrections, and the courts illustrate the ways in which automated tools can reproduce the same biases that plague human decision-makers in these domains (Duncan 2022; Einarsson et al. 2023; Johndrow and Lum 2019).

Democratic governance is predicated on the notion that citizens delegate authority to the state, and that government officials are ultimately agents of, and should be responsive to, the public. Police officers are armed agents of the state, and they operate with enormous discretion and interact frequently with the public (Lipsky 1980). As Soss and Weaver (2017) succinctly put it, "police are our government." As police officers are delegated the state's monopoly on the legitimate use of force to conduct their job, democratic governance requires that they act, and are seen as acting, in ways that are legitimate and accountable to the public.

We lay out a critical role for multidisciplinary collaboration in establishing the measurement framework and building the training data necessary to develop ethical AI models capable of improving government performance and facilitating accountability to a diverse public. First, to enhance democratic accountability and transparency, public preferences must be incorporated into the development of AI tools. This grounding requires the type of survey, interview, and focus group research at which social scientists excel – aggregating and distilling preferences across a potentially heterogeneous set of constituents.

Second, AI models learn from examples that have been labeled or annotated by humans. The quality and accuracy of an AI model's results depend heavily on the human-labeled examples used during training. This is particularly relevant in models trained to analyze politically fraught domains, like hate-speech detection or police-public interactions (Sap et al. 2022, 2019; Golazizian et al. 2024). If labels are created by annotators with a narrow range of perspectives – e.g., only law enforcement officers, or only a pool of professional labelers with similar backgrounds – then those are the only perspectives that will be reflected by the models (e.g., Ouyang et al. 2022).

Given that the assessment of interactions with government officials can be highly subjective (Sierra-Arevalo et al. 2025), tools built to facilitate government accountability must be trained on data reflecting the diversity of perspectives present among the governed. In developing tools to evaluate BWC recordings of traffic stops, it is important  to incorporate perspectives from a diverse citizenry both because their views on these interactions may vary, and because the legitimacy of AI-based government accountability tools relies on capturing the preferences and perspectives of the governed. The human annotation process should capture diverse perspectives regarding the alignment between government-public interactions and identified measures.

Policing is a particularly challenging, high-stakes domain of government transparency and accountability – important in its own right and useful to illustrate the role of social scientists in building effective, community-informed AI tools to enhance government accountability more broadly. We describe how our research team has collaborated with the Los Angeles Police Department (LAPD) on a project to develop community-informed AI models to evaluate officer-driver interactions during traffic stops. Building on our team's experience, we suggest principles and best practices with respect to the development of AI models for police accountability. While policing features particularly high-stakes interactions between the government and the governed, we believe the principles we outline are applicable across domains in which the quality of human interaction shapes the quality of governance.

**An Illustrative Project**

To illustrate our proposed multidisciplinary approach to developing AI tools, and the role of social scientists within it, we describe the Everyday Respect research project, which focuses on communication between LAPD officers and drivers, as captured in BWC recordings of traffic stops. This project is characterized by three features we believe are common across domains for which AI tools can be used to enhance transparency and accountability in government: variation in stakeholder perspectives, sensitive or confidential data, and data that is rich yet incomplete.

1). Community stakeholders vary in their views of what qualifies as appropriate conduct by police officers. Diverse perspectives need to be reflected in both the concepts selected for measurement and the training data used to develop AI models.

2). The data being analyzed are sensitive and confidential. Data sharing restrictions must balance the transparency required for effective oversight and democratic governance with the confidentiality required to protect the privacy rights of individuals depicted in the data.

3). The data collected are rich and multimodal (e.g., they include both audio and video information), and yet still do not fully capture the context in which an interaction occurs. For example, they exclude events preceding the beginning of a recording, which may affect whether an action taken by an officer is appropriate.

By describing a specific AI development project, we illustrate why each element of our approach is important and what community-informed development of accountability tools looks like in practice.

**Diverse Stakeholders, Divergent Experiences**

Across both measurement strategy and model training, we propose a multidisciplinary research method focused on understanding the extent to which officer-public interactions align with the needs and preferences of different stakeholders – not simply trying to capture a single, narrow conception of whether it is good or bad.

***A). Who is a Stakeholder?***

While conceptions and implementation of ethical AI vary, they are mostly focused on consumer AI applications (Adanyin 2024; Farooq, Ramzan, and Yen 2025; Ryan et al. 2024). The concept of stakeholder-informed design usually equates “stakeholders” with “end-users” or “customers.” In the context of AI and BWC footage, the standard approach treats the customers and users as police departments. However, in the context of democratic governance citizens delegate authority to government officials and are subject to the performance of those officials. The goal of AI tools in these contexts is to facilitate the accountability of government agents to the public, making the public the most important stakeholder.[1]

Police departments have the budget and discretion to purchase the AI tools, creating economic incentives for AI companies to be responsive to law enforcement priorities. In contrast, the voices of non-user, non-purchaser stakeholders are often excluded from design processes.[2] Developers have less incentive to incorporate the views of the public – even though we are all deeply affected by the design and implementation choices that are made.

Government stakeholders also often control the data on which AI tools can be trained. In the case of policing, BWC footage, as well as data on crime, enforcement, and police personnel, are often collected, owned and controlled by police departments themselves (Goel et al. 2017; Goodwin 2021). Thus, any actor who seeks to train a new AI tool must do so with the permission of government partners who may end up with *de facto* veto power over the type of work that can be done.

The Everyday Respect team has been able to engage a broad range of non-user, non-customer stakeholders because we are funded by nonprofit and government resources (e.g., the National Science Foundation, Arnold Ventures, and the Microsoft Justice Reform Initiative), and because our team has a mandate from the Los Angeles Board of Police Commissioners – an oversight body – and buy-in from LAPD leadership.

### *B). A Community Informed Measurement Strategy*

A community-informed approach to measurement begins with some combination of survey, interview, and focus group research to assess the perspectives of various groups.[3] This research serves two purposes. First, it identifies the issues most salient to stakeholders. Second, this research facilitates understanding whether and how groups vary in their perspectives.

*Community Informed Measurement in Practice: LAPD Traffic Stops*

In the Everyday Respect project, we used three approaches to understand the preferences of both police officers and members of the public. First, we conducted survey, interview, and focus group research with residents and community organizations from across the city of Los Angeles to assess a broad array of community views on communication with the police. Second, we assessed current LAPD training materials and policy documentation, attended an LAPD traffic stop training at the Academy, and interviewed LAPD leadership to understand the training and policy that guides LAPD officers.Third, we conducted interviews and focus groups with current and former police officers to assess their views on optimal stop conduct by officers as well as the factors that shape the risk environment officers perceive during stops. This allowed us to develop measures that capture the driver behaviors and contextual factors that cause officers to feel threatened or are otherwise likely to predict stop escalation.

Three of the findings from these community assessments illustrate how these assessments shape measurement strategy. First, we identified contested and changing policy areas for specific evaluation. In particular, we focus on the extent to which officers engage in investigative questioning, which is both subject to recent California legislation, and was frequently mentioned in community members' descriptions of negative traffic stop experiences. Nonwhite community members in particular mentioned being asked questions that they felt implied they did not belong in the neighborhood in which they were pulled over or were suspected of criminal activity. In response to these findings, our BWC annotation guidelines are designed to measure investigative questioning. We focus both on a question subject to recent California legislation, "Do you know why I pulled you over?"[4], as well as questions about parole and probation and questions that may imply a driver does not belong in a particular area. We also annotate items specific to evaluating officer's compliance with a new pretextual stop policy adopted by the LAPD in March of 2022.[5]

Second, racial disparities in search behavior by the LAPD during traffic stops were reported in the *LA Times* and elsewhere in the leadup to our study, and were top-of-mind for the community organizations represented in our focus groups.[6] We thus annotate in detail not only who is searched but also whether and how consent for searches is solicited. By measuring search behavior on the basis of BWC recordings, we can capture both the process by which consent is or isn't obtained and directly compare the BWC record of searches and seizures with the officer-generated administrative record of the same stop. This last piece is critical because research and policy reform is built on the assumption that administrative data is accurate, even if we have evidence of its inaccuracy (Eckhouse 2022; Poston and Rubin 2014).

Third, across all races of survey respondents, we find a shared priority on receiving respectful treatment from police officers. Thus, “respect” is the subjective construct our team measures in the most detail. **Figure 1** provides descriptive support for this focus, illustrating the proportion of topics and themes identified through natural language processing (BERTopic modeling) of responses from a large-scale survey we conducted with Los Angeles residents.[7] In this survey, participants were asked to describe both the positive and negative behaviors and communication styles they preferred or sought to avoid from police officers during traffic stops. Across all racial groups, respondents not only referred repeatedly to the concept of respect, but also offered similar descriptions of what respect looked like. Respondents associated respect with professionalism, transparency, kindness, and sympathy; whereas aggressiveness, condescension, racial insensitivity, and lack of empathy were widely regarded as indicators of disrespect and behaviors they wished to avoid experiencing.

**Figure 1. Topic proportions from a large-n survey on public expectations of police behavior during traffic stops**

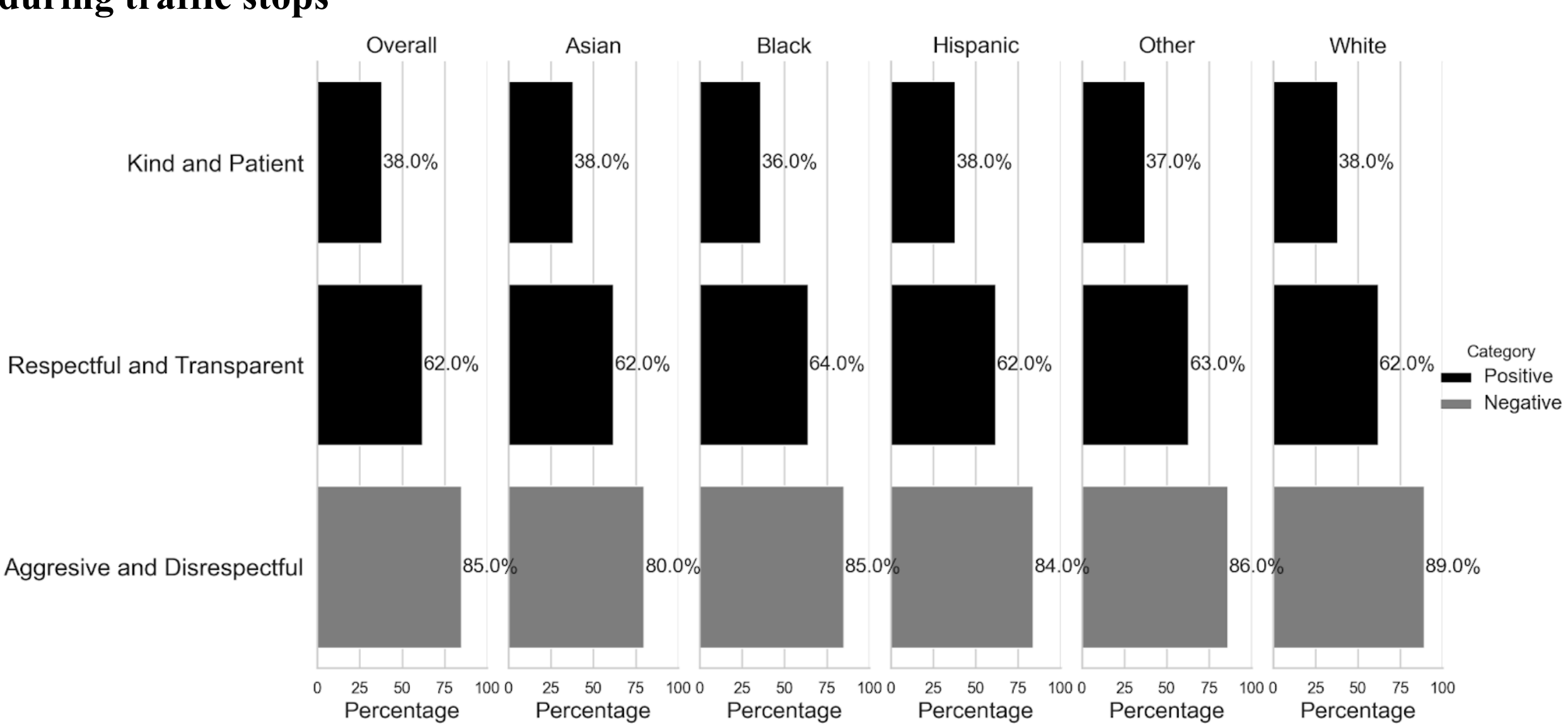

Note: Several of the model-generated topics from each category were removed, as they did not relate to preferences for traffic stop communication and actions. The pool of adult survey respondents was weighted to Los Angeles City population benchmarks derived from U.S. Census data for age, race, education, and income.

In contrast to the broad agreement across respondents in their preference for–and descriptions of–respectful interactions, our interview data revealed racial differences in *why* respect is considered important. As illustrated by the quotes in Table 1, Black and Latine respondents perceive negative and violent outcomes from traffic stops to be much more likely than White respondents do. For Black and Latine drivers, respect signals that a stop is likely to be lawful and peaceful, alleviating some of the fear with which they may enter the interaction. White drivers, in contrast, are less likely to perceive violence as a possibility, fearing instead outcomes such as an expensive ticket. Thus, for White drivers, respect does not play the same role in reducing the fear of violence or unlawful treatment and is more often valued simply as a feature of a more pleasant interaction.[8]

**Table 1** provides additional insight into these racial differences from our interviews, highlighting how respondents describe their experiences and fears regarding police encounters. Black and Latine respondents frequently link traffic stops to feelings of fear and potential escalation, with some expressing concern that a simple interaction with police could have fatal consequences. A Black male respondent described the stress of navigating a stop, worried that a single misstep could *"make the situation so much worse."* A Latine female respondent similarly articulated a growing fear of traffic stops, framing them as moments where drivers now worry *"for their life."* By contrast, a White female respondent did not associate police stops with the same level of physical danger. Instead, as reflected in the quote below, she described her concerns in terms of financial or legal repercussions, such as receiving a ticket or points on her license. One White male respondent also reflected on how his race shaped his past interactions with police, recognizing that his White identity made him *"less likely to be pulled over"* compared to his Black and Asian friends.

Together, these responses reinforce the idea that while respect is universally valued, its significance differs across racial groups. For Black and Latine respondents, respect from police officers signals that an interaction is less likely to escalate into violence. For White respondents, it remains important but is framed more as a matter of fairness and professional courtesy rather than personal safety.

**Table 1. Differing Perspectives on Traffic Stops**

| Quote | Demographic |
|---|---|
| "But now, there is no peace, everything ends up in an escalation, [...] everything ends up with the potential for fatality." | Black Female (59) |
| "Like I don't know what I should say here like cause I'm like stressed. It's – I could say something that would like make the situation [of being pulled over] so much worse for me." | Black Male (28) |
| "From the driver's perspective. Yeah, like [the feeling of being pulled over] now, I mean, there's more of a fear, a fear for their life." | Latine Female (33) |
| "I think I felt scared about like, oh my god, am I gonna get a huge ticket, or am I going to get points on my license? Something like that. | White Female (27) |
| "[M]y friends, who were mostly Black and Asian [...] were always like, 'Yeah, you, you drive'. You know, when we would go places. So I was always the driver, and I didn't fully realize it at the time, but I get it now. You know it was because, I, I, I'm White, so I was less likely to be pulled over." | White Male (38) |

As a result of the interviews, we designed our annotations of the BWC footage to measure fear as expressed by both officers and drivers at the beginning, middle, and end of each stop. We can use these data to test the extent to which respectful conduct from officers and drivers (as perceived from a range of annotator perspectives) shapes interactions between drivers and officers.

Stakeholder research also informs data annotator recruitment. As discussed in the following section, we focused our recruitment of annotators on ensuring broad representation of different Los Angeles communities among our annotators, with particular attention to recruiting individuals who have been arrested or incarcerated in the past and former police officers.

### ***C). Multiperspective Training Data for Multiperspective Models***

Traditional AI or machine learning analyses of interpersonal communication typically assume a single ground truth—an objectively "correct" or unbiased interpretation—such as whether a statement is respectful or an action is threatening.[9] In contrast, we argue that reasonable observers can legitimately differ in how they characterize the same interaction or behavior. Thus, a core objective is to develop models that reflect diverse, valid viewpoints, recognizing disagreements between observers not as noise, but as valuable evidence of varied perspectives.

Historically, the development of a single "ground truth" has privileged the perspectives of the powerful (Lee 2018; Barabas et al. 2020). Our approach to measuring communication moves away from accepting single viewpoints rooted in power differentials and instead expands the range of perspectives upon which AI tools are developed.

The assumption of a single ground truth is prevalent in both AI-driven methods and traditional social science approaches.[10] For instance, the dominant social science method for studying police-public interactions, Systematic Social Observation (SSO), aims for "neutral" measurement. Historically, SSO has been applied through in-person observations (Brunton-Smith 2018; McCluskey et al. 2023; ~~J. D.~~ McCluskey and Reisig 2017) and more recently through human annotation of body-worn camera recordings (Chillar, Pizza, and Sytsma 2021; Nassauer and Legewie 2021; Worden et al. 2025). In AI and machine learning contexts, the single-ground-truth framework (often, a singular "majority opinion") similarly remains widespread. However, a growing number of researchers now advocate for annotation and modeling practices that capture and incorporate multiple valid perspectives (Kapania, Wang, and Taylor 2023; Muscato et al. 2024; Wang et al. 2024; Chochlakis et al. 2025).

Techniques for training AI models to handle multiple, divergent interpretations of subjective tasks are rapidly advancing (Akhtar, Basile, and Patti 2020; Kanclerz et al. 2022; Davani, Aida, and Prabhakaran 2022; Fornaciari et al. 2021; Hejabi et al. 2024). These emerging approaches fundamentally embrace disagreement among human annotators not as statistical noise, but as meaningful reflections of genuine ambiguity and variation in human perception (Akhtar et al. 2019; Plank et al. 2014; Prabhakaran et al. 2021).

Our approach to multi-perspective modeling operates in two stages. In the first stage, data are collected from a small pool of annotators. These data serve as a foundation for building a multitask model that captures the similarities and differences in annotator perspective. Informed by the first-stage annotations, in the second stage we recruit new annotators that we expect to feature the disagreements identified in the first stage. This allows us to focus our resources on the types of annotators that provide maximum information about distinct perspectives. Empirical evaluation of model improvements from annotator diversification suggests it is a valuable tool for improving model performance (Golazizian et al. 2024; Muscato et al. 2024; Wang et al. 2024). The human annotations are then used to train AI tools to recognize those same constructs.

We propose three general principles for recruiting and training diverse annotators for subjective annotation tasks. 1) Targeted recruitment of annotators who vary across the dimensions most salient to the domain in question (e.g., recruiting both former police officers and people with a history of past arrests). 2) Annotator training that establishes a shared understanding of concepts but leaves space for individual perspectives on contested constructs. 3) An annotator-in-the-loop process to incorporate diverse annotator perspectives into refined item definitions and training guidance. The following section illustrates those principles in practice.

*Multiperspective Annotation In Practice: LAPD Traffic Stops*

To create training data for models of police traffic stops, we hired human annotators to watch BWC recordings associated with a stratified random sample of 1000 LAPD traffic stops, which consists of about 1800 BWC recordings because of multiple recordings for many stops. We recruited annotators that we expect to have varied and contrasting perspectives on police-public interactions. In practice, this entailed recruiting annotators with variation across race, gender, age, and police-related experiences, investing particular effort to recruit people with a history of past arrests or incarceration and people who have worked in law enforcement. Partnerships with organizations like HomeBoy Industries[12] facilitated recruiting individuals with past arrests and/or gang affiliations, while annotators with law enforcement backgrounds were identified via referrals from the LAPD and USC's Safe Communities Institute.

Training a diverse group to annotate subjective and emotionally/politically charged items requires balancing consistent variable definitions with preserving diverse perspectives and avoiding enforcement of a "professional vision"(Goodwin 1994). With concepts like respect or fear, we seek to establish a shared understanding of how items are defined but allow annotations to reflect variation across individuals in whether they believe a given behavior meets that definition.

We provide an annotation guide detailing definitions and examples of both objective and subjective variables to provide a baseline understanding of the concepts to be annotated.[11] In an initial training session, annotators review these definitions, practice on public videos, and are encouraged to provide their own perspectives and feedback. After this, we conducted small group training sessions in which we grouped annotators with similar backgrounds to encourage honest, open discussion of contentious topics. We did this to make it less likely that people with different perspectives would try to convince each other whose perspective(s) was right, which would partially defeat the purpose of multi-perspective annotation.

Unlike traditional methods, where annotator training concludes after onboarding, we implement a continuous "annotator-in-the-loop" process to enhance data quality and annotator well-being (Schmer-Galunder et al. 2024; Diaz et al. 2022; Carrión and Casacuberta 2022). For an annotator's initial weeks, we conduct weekly check-ins to address concerns, refine processes, and monitor well-being. These sessions identify technical issues (e.g., corrupted video files), conceptual challenges (e.g., ambiguous variables), and human factors (e.g., emotional fatigue).

Incorporating this feedback into the annotation process enables adaptive improvements, supporting annotator engagement and the production of high-quality, multiperspective data.

The annotation process we follow is labor-intensive and costly. However, when a preeminent goal is to enhance democratic accountability, it is essential to implement an annotation process that incorporates the diversity of perspectives present in society.

*Perspectives Vary by Annotator*

Survey and interview data show that annotators from different backgrounds enter the annotation process with different perspectives on policing (Sierra-Arévalo et al. 2025). Do those perspectives shape how annotators evaluate subjective features of traffic stop interactions?

During our annotation process, we selected 295 videos[13] for which we collected 2-3 annotations per video of subjective items – a total of 813 annotations. We focus here on five items. *Danger: Driver* and *Danger: Officer* capture assessments of the following question: "How in danger or threatened do you think the Driver/Officer felt during this traffic stop?" *Respect: Driver* and *Respect: Officer* assess the overall level of respect shown by the driver and the primary officer during the interaction and *Stop: Legitimacy* captures the annotator's overall assessment of stop legitimacy. All outcomes are measured as 5-point Likert scales with a neutral midpoint.

To demonstrate the importance of incorporating diverse perspectives for subjective items, we also analyze the variation in how annotators assess two objective items, *Any Person Searched?* and *Any Contraband Found?,* which capture whether a given annotation reports any officers conducting a search of a person or vehicle during the stop and whether, conditional on a search, they find any contraband (e.g., guns or drugs). These are drawn from another 903 videos of 2-3 annotations, resulting in an additional 2,088 annotations.

With multiple annotations per video, we can explore the extent to which variation in our measures of these items is explained by characteristics of the video or characteristics of the annotator. For the items measured on a Likert scale, we decompose the variance for each item using a Linear Mixed-Effects Model estimated via Restricted Maximum Likelihood. We model **Video-** and **Annotator-level** variances as crossed random effects. This approach allows for the decomposition of total variance into three components: variance attributable to differences between videos, variance attributable to differences between annotators, and the residual variance (i.e., the variance left to be explained). The proportion of variance for each component is calculated as the ratio of the respective estimated variance component to the total sum of estimated variance components. We model the binary objective outcomes similarly, using a Logistic Mixed-Effects Model, and following a latent variable interpretation of the logistic regression, fix the residual variance to the variance of the standard logistic distribution (Devine et al. 2024).

Figure 2 shows that, for relatively objective items that capture whether or not an event occurred during a stop, variation across annotators is low (see the two rightmost columns). This low cross-annotator variance suggests that heterogeneity in annotator characteristics is likely to be relatively unimportant in shaping model performance for (mostly) objective items. In contrast, for subjective items the variance across annotators is substantial – in some cases as large or larger than the variation across videos.

***Figure 2: Variation Across Annotators and Videos***

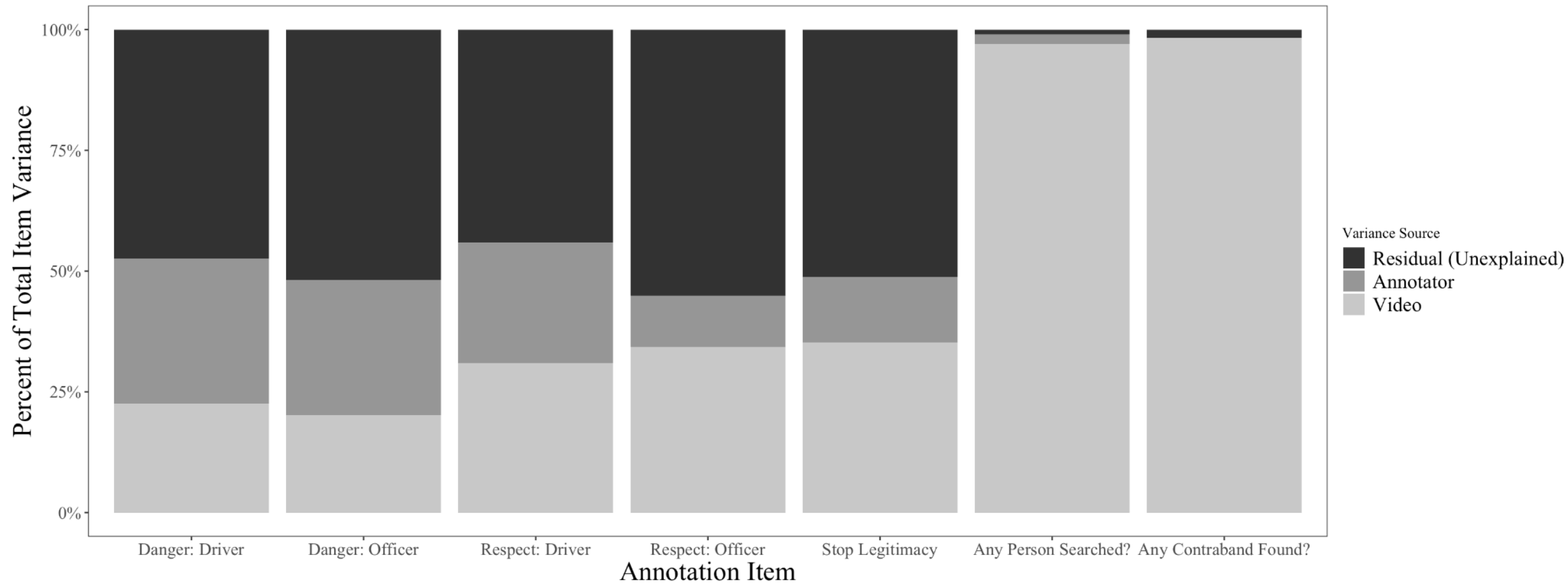

For items such as *Danger* and *Respect*, the substantial proportion of variance attributable to the **Annotator** suggests that annotators' perceptions are significantly influenced by their internal baseline or dispositional traits. This indicates that the rating, at least to some degree, reflects the rater's own psychological or cultural lens. In these instances, the "eye of the beholder" contributes as much to the measurement as the stimulus itself.

To further demonstrate the importance of annotator perspectives, we present the ratio of annotator-video variances in Figure 3. On items of critical importance to police departments and social scientists alike, who evaluates the video is almost as, or more important than, the content of the video itself in explaining the variance in the outcome measures that are generated.

***Figure 3: Ratio of Annotator-Video Variance***

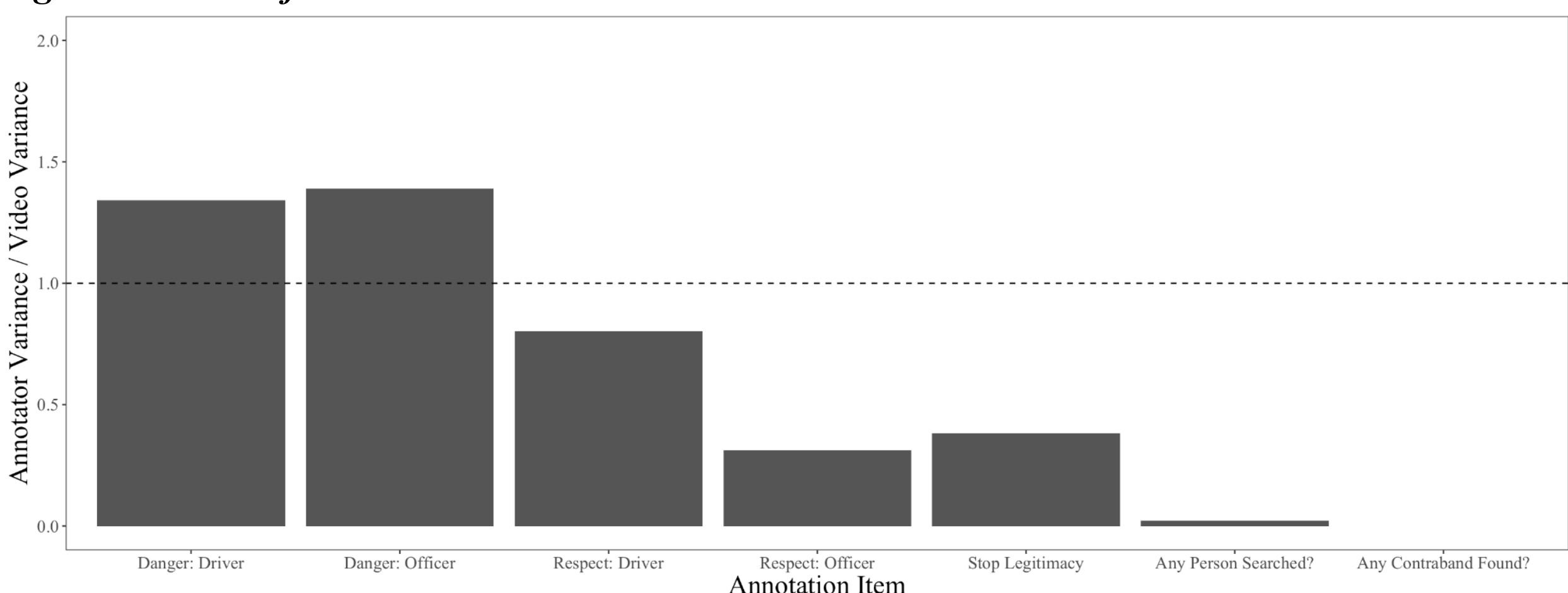

These results confirm the relevance of incorporating diverse perspectives into AI training data: People with different experiences and characteristics vary in their perceptions of the subjective features of government-public interactions. When modeling subjective outcomes in contexts like government accountability in which variation in perspectives is important, this suggests there is value to recruiting annotators from a range of backgrounds and modeling annotator variance as a function of annotator characteristics. Only by capturing diverse perspectives at the annotation/labeling stage can one train AI models that include the perspectives of the governed. If AI models are trained on a narrow range of perspectives, there is significant risk that their use in governance tasks unintentionally amplifies the same biases and power imbalances that afflict modern democratic politics (e.g., Turner Lee 2018; Barabas, Colin Doyle, and Dinakar 2020).

**The Advantage of Using AI Tools to Analyze Sensitive Data**

Video and audio footage, such as that captured by BWCs, is a particularly valuable source of data because it captures details of government-public interactions that are not included in other official records, providing a less-mediated record of events (Knox, Lowe, and Mummolo 2020). However, the same richness that makes video data so valuable also introduces privacy-related concerns. In the traffic stop context, BWC data typically include identifiable characteristics of the civilian(s) being stopped.

Government-private citizen interactions captured on BWCs are a recording of a conversation. Generally, the legality of recording has two parts, the situation where the conversation occurs and the consent or awareness of the recording by one or more people present. Even if law enforcement or other government agencies have the right to record interactions, there are complexities around using the recordings for research or public records, because many recordings are made in an environment where people may have an expectation of privacy under the Fourth Amendment (Cal. Penal Code § 632(a) 2021). California, for example, requires all parties present to consent to a recording if the participants have a reasonable expectation of privacy, such as within a private home (Code 1967). There is a challenge inherent in (1) protecting the privacy of members of the public who interact with government officials and (2) making data about these interactions publicly available to monitor government officials' actions.

Machine learning and AI models provide an opportunity to study high-stakes interactions in detail, while also protecting the rights of the individuals depicted in the data. Because these models can analyze recordings of interactions without a human viewing the video, the models can reveal patterns in the behavior of public servants without violating the privacy of the people depicted in the data.

**The Role of Multidisciplinary Collaboration: A Set of Best Practices**

The development of AI models, even for the purpose of government accountability, is generally considered the *de facto* domain of engineering and computer science. Decades of work in social science, however, demonstrate that political preferences and experiences with governance vary across groups. Our work on policing further suggests that, despite groups having a uniform desire for respectful treatment from officers, they often have widely divergent reasons for this preference, and different concerns with regard to what may occur if respect is absent. When viewing the same interactions, annotators from different backgrounds may vary systematically in their perceptions. If the preferences and perspectives of the governed are not reflected in how AI models are developed, it becomes less likely that the application of such models will effectively support the performance and accountability of democratic governance.

Thus, social scientists and humanities scholars – political scientists, sociologists, economists, anthropologists, psychologists, and others – have a role to play in ensuring that AI models are developed in a way that accounts for the experiences and perspectives of diverse stakeholders. If government legitimacy rests on democratic principles, then so too should the models that we use to evaluate government action. Moreover, the task of understanding diverse perspectives is well-suited to the methodological training of social scientific researchers. To facilitate this effort and abstract away from the Everyday Respect project, we propose a set of best practices for building community-informed AI models that advance government transparency and accountability. We summarize our approach in Figure 4.

**Figure 4: An Approach to Developing Community-Informed AI Models**

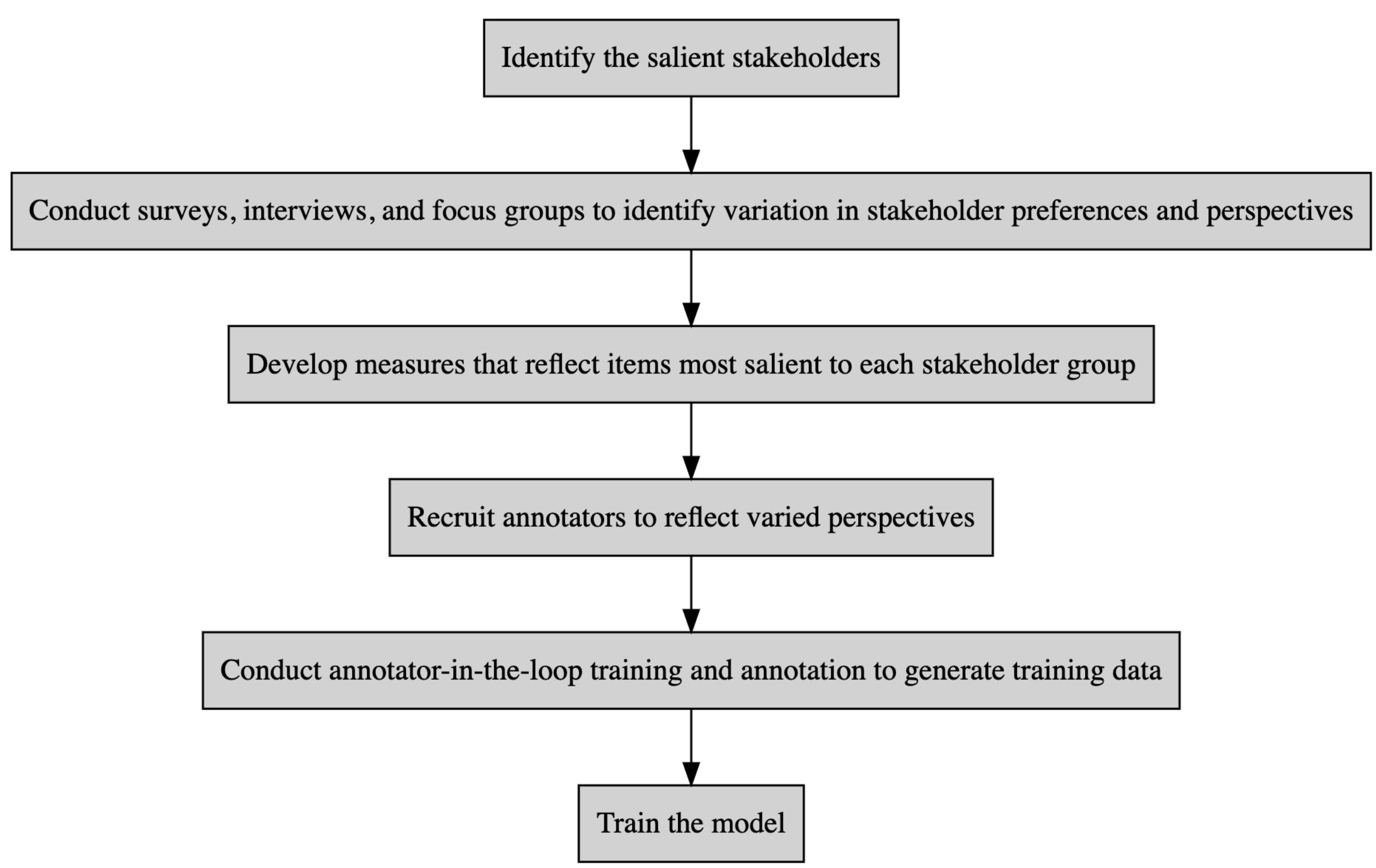

First, researchers must identify the salient stakeholders within a given domain. For models designed to enhance democratic accountability, stakeholders include not only government agencies and their employees but also their democratic principals—citizens or constituents. Researchers should recognize that even within stakeholder groups (e.g., bureaucrats, citizens), preferences and experiences are likely to vary substantially.

Second, researchers should use surveys, interviews, or focus groups with representative samples from each stakeholder group to identify key interaction features and understand how stakeholders define and interpret them.

Third, insights into stakeholders' divergent preferences and experiences should directly inform the measurement strategy, ensuring that it encompasses the elements most salient to each group. Researchers should create an annotation manual that captures both objective (where high agreement between annotators is expected) and subjective dimensions (where systematic disagreement between annotators is expected) of the phenomena under study.

Fourth, researchers should conduct targeted recruitment of annotators to ensure variation across salient characteristics (e.g., relationship to or past experience with the agency being studied). While a larger and more diverse sample is always preferred, researchers should use information from the stakeholder research to identify which characteristics are most important to target in annotator recruitment to ensure variation across the dimensions most likely to relate to variation in perspective.

Fifth, researchers should adopt annotator-in-the loop training, continuously refining the annotation manual in response to annotator feedback. The goal is to establish a shared understanding of concepts, while explicitly allowing for variation in perspectives. In order to create a multiperspective training dataset, multiple annotators should evaluate each interaction.

Finally, these multiperspective annotations should be deployed in a modeling framework that makes use of the information present in annotators' disagreements, rather than discarding this information as noise in an attempt to estimate a single ground truth.

Taken together, these steps make it possible to develop multiperspective AI models that reflect the preferences and viewpoints of the governed. These models can then measure government-public interactions at scale and at speed. Once models are trained, the cost of analyzing additional data shrinks, supporting a wide-range of applications, from scientific research to public transparency to real-time monitoring, supervision, and coaching of police officers and other government officials.

**Limitations and Challenges**

The project we describe illustrates an inductive approach to developing community-informed, multiperspective models. Our approach integrates social science knowledge into the development of AI models in a way that is particularly valuable with respect to police and government accountability. However, the approach we advocate retains limitations that we hope future methodological innovations can address.

Perhaps most notably, while recruitment of annotators can be targeted, budget constraints push developers toward annotator pools that are non-randomly selected and medium-n. It likely requires hundreds, rather than dozens, of annotators to capture this diversity fully. Targeted recruitment makes the most efficient use of a limited number of annotators, but larger scale annotations would be strongly preferable.

Second, even with only a medium-n annotator pool, the approach we promote is time and resource-intensive. Annotating BWC recordings of 1000 traffic stops using this method has cost roughly $600,000 in annotator wages, in addition to thousands of hours of researcher time to develop the measurement strategy, customized annotation software, design and implement the modeling, and manage the relationships with our many stakeholders. Thus, while we are able to clearly articulate the role and value of political scientists and other social scientists in the development of AI models, it is not the case that it is easy for an individual academic to launch a project similar to the one we describe. It is challenging to build expensive projects that require large multidisciplinary teams and deep collaboration with government agencies.

Third, while advances in AI, along with our approach, facilitate the development of community-informed AI models, governments themselves often control access to the data necessary to build these models. Our team has gained unique access to sensitive data and works with the LAPD closely, other teams of researchers may not have the same experience. Even if they do, the government agency they work with may retain "veto power" over what content they access, and potentially, the development of measurement strategies and annotation manuals. Approaches to building community-informed AI models may then be stymied by the government one wishes to assess.

These limitations are significant and imply that efforts to build AI tools to enhance accountability and transparency require a large, well-financed research team. It is important not to trivialize the challenges of building models that recognize and account for the importance of the multiple perspectives relevant to democratic oversight. At the same time, if social scientists fail to participate in building these tools it essentially abdicates their development to groups without the expertise, or without the incentives, necessary to create tools that can improve democratic governance.

**Conclusion**

In-person interactions between government officials and members of the public are central to high-quality governance. These interactions are increasingly captured on video—at government offices, at passport control, or via body-worn cameras (BWCs) used by law enforcement officers. These data have not been leveraged in large-scale efforts to enhance democratic accountability, partly due to the challenge of reviewing and learning from vast amounts of footage. However, rapidly advancing AI technology enables the analysis of rich audio, video, and transcript data, potentially enhancing our understanding of these interactions and increasing both transparency and accountability in government-public interactions. To realize this potential, AI models need to be designed in accordance with principles of democratic accountability and with the ability to account for the varied preferences and perspectives of democratic stakeholders.

We describe an approach to address the theoretical and practical challenges of accounting for such varied perspectives in a particularly challenging domain – urban policing. Police officers have a significant effect on public attitudes towards government and have tremendous discretion in how to perform their jobs. Police departments are also often the single biggest line item in municipal budgets. Given their social, legal, and financial importance, they are an essential element of governance to understand, and significant variation in people’s experiences with and perceptions of police make it a domain where accounting for multiple perspectives is critical.

To date, the AI-development community has not established a set of best practices for building tools for government accountability. The process we outline focuses on democratic accountability/oversight and the related importance of ensuring that diverse perspectives are understood and incorporated into the development of the AI models. We believe much of this approach is relevant for any context in which a democratic government is an important actor, because in such settings it will be critical to consider the diverse preferences and perspectives of the governed. These new technologies have tremendous promise to facilitate democratic accountability and transparency, but only if these concerns are forefront in the development of the AI tools.

The development and implementation of these models is neither hypothetical nor speculative. With respect to policing in particular, AI models are already on the market that analyze BWC footage to evaluate officer policy compliance and even to write the rough draft of police reports (Watts, White, and Malm 2025; Adams, McLean, and Alpert 2024; Heydari et al. 2024). However, commercially-developed tools are unlikely to be developed in a manner that reflects the perspectives and preferences of diverse constituencies; these tools in many cases are designed to serve the needs of government agencies, not the needs of the public. The role of social scientists is to help develop and implement better tools to achieve better outcomes.

The approach we outline focuses on the role of political scientists in constructing multiperspective models to facilitate government accountability. However, society increasingly has access to rich data on a wide variety of high-stakes human interactions, including those between teachers and students, physicians and patients, or managers and employees. In any given domain, there are different sets of stakeholders, different sources of variation in preferences and perspectives, and different power dynamics and privacy concerns. Nevertheless, we believe that both the broad principles we lay out here for multiperspective model development, and the essential role of social science in the application of those principles, likely remain valid across a range of human interactions. We hope that political scientists will engage deeply with the development of AI models for government accountability, and that the practice of deep social science engagement in model development extends more broadly across fields.

**Endnotes**

1. In its Blueprint for an AI Bill of Rights, the U.S. Office of Science and Technology Policy writes that "[AI] systems should be used and designed in an equitable way." We take this argument a step further to argue that, in the context of tools for government accountability, they must be used and designed in a manner consistent with democratic principles.

2. Civil liberties advocates have long criticized lack of community input into decisions regarding the deployment of surveillance technologies, as well as the use of non-disclosure agreements and other tools to limit public knowledge about when and where certain technologies are deployed (e.g., Joh 2017; ACLU 2016)

3. In contrast to community-led research, in community-informed research it is researchers, rather than community members, who make final design decisions (Bammer 2019; Ray and Miller 2017).

4. Assembly Bill 2773 requires officers to state to the driver/pedestrian the reason for the stop before engaging in questioning them, as well as include in their official stop report the reason why a driver was pulled over. The legislation passed in 2022 and took effect on January 1, 2024.

5. The 2022 LAPD policy affecting pretextual stops is described at (Los Angeles Police Department 2022).

6. The *Los Angeles Time*s article discusses the higher stop rates, but lower rates of identifying contraband among Black and Latino drivers (Potson and Chen 2019).

7. We conducted a two-wave survey of approximately 2,000 Los Angeles residents in November 2023 and June 2024 to assess public preferences regarding officer conduct during traffic stops, personal experiences with such stops, and overall perceptions of police performance and legitimacy. To ensure representativeness, we applied survey weights to align the sample's demographic composition with that of Los Angeles in terms of age, race, education, and income. For further details, see Sierra-Arévalo et al. (2025). BERTopic (Grootendorst 2022) leverages an underlying Bidirectional Encoder Representations from Transformers (BERT) (Devlin et al. 2019) text-embedding model, clustering techniques, and a class-based variation of the Term Frequency-Inverse Document Frequency (TF-IDF) measure to identify coherent topic representations from text data.

8. For a full description of these results, see Sierra-Arévalo et al (2025).

9. For a review of standard approaches to minimizing "bias" in human annotation, see (Kapania, Wang, and Taylor 2023).

10. The single-ground truth approach is adopted in even the best current work using machine learning methods to analyze BWC footage (e.g., Camp et al. 2024).

11. The annotation manual and all training materials are available online at https://osf.io/py6md/ and the custom multiperspective annotation software we developed is available open-source (Hejabi et al. 2025).

12. Homeboy Industries is a gang intervention, rehabilitation, and re-entry program based in Los Angeles.

13. For additional information on these videos, the annotators, and the items annotated, see Appendix B.

**Data Availability**

Deidentified versions of the survey responses and annotation data analyzed during the current study, along with replication code, are available via Open Science Framework (https://osf.io/py6md/). The bodyworn camera recordings on which the annotations are based are not publicly available, per the terms of the Professionals Services Agreement in place between the research team and the City of Los Angeles. The text of this professional services agreement is available from the corresponding author on reasonable request.

**Acknowledgements**

The authors would like to thank the community organizations and individuals who participated in focus groups and interviews for this project, as well as the Los Angeles Police Department, the Los Angeles Police Commission, and the City of Los Angeles for their cooperation and support. Thanks to all of the video annotators, on whose work the entire project rests, to the undergraduate research assistants in the SPEC Lab and to Oliver Chavez, Sarah Wilson, and Kerby Bennett for their project management. Thanks also to our funders: the Microsoft Justice Reform Initiative, the National Science Foundation (CIVIC program and Smart & Connected Communities Program), Google (Award for Inclusion Research), Arnold Ventures, and USC (Zumberge Interdisciplinary Research Award).

**Competing Interests**

Dr. Narayanan is Chief Scientist and Co-Founder with equity stake of Behavioral Signals, a technology company focused on creating technologies for emotional and behavioral machine intelligence in consumer services. He is Chief Engineering Science Officer and Co-Founder with equity stake of Lyssn.io a technology company focused on tools to support training, supervision, and quality assurance of psychotherapy and counseling. He is also a Visiting Faculty Researcher at Google, LLC.

Georgios Chochlakis was an Applied Scientist Intern at Amazon Web Services, Inc. during some of the time he was working on this project.

**Ethical Approval Statement & Informed Consent**

Approval for the human subjects research portions of this study was granted by the Institutional Review Boards at the University of Southern California and the University of California, Riverside. All research pertaining to this project was conducted in accordance with ethical research standards as laid out in the 1964 Helsinki Declaration and the 1979 Belmont Report and with the protocols approved by the Institutional Review Boards (IRB) at the University of Southern California (USC) and the University of California, Riverside (UCR). Approval for the interview and focus group portion of this research was granted by the USC IRB on May 17, 2022, as study number UP-22-00362. Approval for the study of variations in annotator perspectives was granted by the USC IRB on February 21, 2024, as study number UP-24-00018.

Approval for the survey portion of this research was granted by the UCR IRB on September 8, 2022, as study number HS-22-116..

All three elements of this study were categorized by the IRB at the relevant Principal Investigator's home institution as "Exempt" category studies. As such, participants were provided with information about anonymization practices and the risks and benefits associated with participation, but signed informed consent forms were not gathered. Informed consent was obtained from survey respondents, interview respondents, focus group participants, and video annotators immediately prior to each individual's participation in the project. For interview and focus group participants, consent was obtained by Dr. Lauren Brown, Dr. Brittany Friedman, Raquel Delerme,  and Dr. Benjamin Graham between May 1, 2022 and December 1, 2023. Consent for survey research was obtained in two waves by the vendor Prime Panels by Cloud Research in November 2023 and June 2024, under the supervision of Dr. Nicholas Weller. For video annotators, informed consent was obtained by Oliver Chavez and Dr. Graham between February 1, 2024 and June 1, 2025, depending on the start date of the annotator.

**Online Appendix for:**
**Community-Informed AI Models for Police Accountability**

## Appendix A: Large-N Survey Details

To understand the preferences of Los Angeles residents about desired and undesired police behavior during traffic stops, we drew on a survey of approximately 2,000 respondents. Our survey was fielded in two waves during November 2023 and June 2024 through PrimePanels by Cloud Research, which recruited respondents from a provided list of ZIP Codes in Los Angeles proper. While we did not draw a representative sample of residents for this survey, we weighted our completed survey sample to match the demographics of Los Angeles based on age, race, education, and income from the U.S. Census (American Community Survey 2022). After preprocessing responses, we were left with 1,141 respondents from the November wave and 917 respondents from the June wave for a total of 2,058 after merging the two waves. To better assess preferences for policing behavior, we formatted our survey questions with the following open-ended prompts:

> *For the next two questions, we would like you to think about an interaction between you and a police officer during a normal traffic stop. For example, imagine you have been pulled over for speeding on a local road. Please answer the following questions about communication with the officer during the interaction.*

> *Please describe how you would [NOT] like the officer to communicate with you. Feel free to include specific words or tone you would like the officer to use.*

We followed up the question about communication with a question asking respondents to:

> *Please describe any actions you would [NOT] like the officer to take during the traffic stop.*

After obtaining our survey responses, we performed a series of preprocessing steps to ensure consistency and reliability in our data. This process consisted of (1) concatenating verbal and action responses for positive and negative prompts, resulting in two primary categories: positive and negative preferences; (2) removing all rows with missing data across any response categories, ensuring that only complete cases were included in the analysis; and (3) excluding respondents under the age of 18 and those who marked that they did not reside within Los

Angeles. These preprocessing steps ensured the dataset was complete, relevant, and focused on the target population, providing a robust foundation for our topic modeling.

**Topic Modeling**

In the field of natural language processing (NLP), topic modeling is essential tool for understanding and organizing large text corpora across the political sciences (e.g. Osnabrügge, Ash, & Morelli 2023; Rodman 2020; Wilkerson & Casas 2017). Unsupervised topic modeling techniques, such as Latent Dirichlet Allocation (LDA) (Blei, Ng, & Jordan 2003) and Non-negative Matrix Factorization (NMF) (Lee & Seung 1999), along with its variations, has been widely used to identify latent topics in unstructured text data. However, these methods often fall short as they disregard the contextual relationships between words in the text (e.g. Egger & Yu 2022). Similarly, text embedding techniques like Top2Vec (Angelov 2020), while overcoming the limitations of traditional topic modeling methods, can rely on a centroid-based assumption that may not always hold in certain clusters of documents.

To overcome these challenges, BERTopic (Grootendorst 2020) leverages an underlying *Bidirectional Encoder Representations from Transformers* (BERT) (Devlin et al. 2019) text-embedding model, clustering techniques, and a class-based variation of the Term Frequency-Inverse Document Frequency (TF-IDF) measure to identify coherent topic representations. Broadly, there are several independent steps that BERTopic follows to generate topics: (1) The embedding of documents using a Sentence Transformer model; (2) Reducing the dimensionality of the produced word embeddings; (3) Clustering of the embedded documents; (4) Pre-processing the corpora via tokenization; and (5) The selection of the most representative documents for each cluster using a custom class-based TF-IDF (Grootendorst 2020).

Text-embedding models transform text documents into numerical vectors, which can then be used to analyze text data in a mathematical space. BERTopic relies on sentence-embedding models that convert sentences of text to capture the semantic meaning of entire sentences instead of relying on the exact tokens found in the text (Reimers & Gurevych 2019). A dimensionality reduction model, like the Uniform Manifold Approximation and Projection (UMAP) algorithm (McInnes, L., Healy, J., & Melville, J. 2018), is applied to the embedded documents to visualize the clusters in a two-dimensional space and reduce the complexity of the embeddings while preserving their essential structure.

A clustering algorithm, like the Hierarchical Density-Based Spatial Clustering of Applications with Noise (HDBSCAN) (Campello, M., Moulavi, D., & Sander, J. 2013), is then used to cluster the embedded two-dimensional documents containing similar semantic content. Once these clusters are formed, BERTopic then identifies the most representative words found in each cluster by analyzing the original text associated with the clustered embeddings. These words can be pre-processed (e.g. tokenized) to increase the topic representation quality before weighting and classifying them.

Finally, BERTopic employs a class-based variant of the traditional TF-IDF, known as c-TF-IDF. Like the traditional TF-IDF, c-TF-IDF is a measure of term importance in a document, but it is specifically designed to highlight what makes documents within a single cluster different from those in other clusters, thereby enhancing the interpretability of the topics. Class-based TF-IDF can be represented as:

$$W_{x,c} = |tf_{x,c}| \times log\ log\ \left(1 + \frac{A}{f_x}\right)$$

where:
- $tf_{x,c}$ is the frequency of the term *x* within the cluster *c*;
- $f_x$ is the frequency of the term x across all clusters; and
- *A* represents the average number of words per cluster.

To this end, we use BERTopic to identify the most salient topics in our survey of Los Angeles residents and use the following combination of submodels to do so: (1) sentence embed our survey responses using the Salesforce Research Embedding (SFR-Embedding-2_R) sentence-transformer model (Meng et al. 2024), (2) apply the UMAP algorithm with cosine similarity for dimensionality reduction, (3) apply the HDBSCAN algorithm for document clustering, (4) use CountVectorizer (Pedregosa et al. (2011)) to pre-process the clustered word responses, and (5) select the most representative survey responses for each cluster using a class-based TF-IDF.

In addition to these steps, we fine-tune our topic representations by taking the following extra steps: (6) integrate OpenAI's GPT-4o to generate suggested topic labels and descriptions for each cluster through prompt engineering, and (7) reduce topic outliers by building a supervised BERTopic model with the most representative topics from the unsupervised model and classifying the remaining outlier topics based on the supervised model's predictions.

**Incorporating Large Language Models (LLMs) with BERTopic Modeling**

To enhance the interpretability of topics, we use BERTopic's compatibility with large language models by integrating OpenAI's GPT-4o to generate concise topic labels and descriptions. Using representative keywords and a sample of documents from each cluster, GPT-4o synthesized short yet meaningful descriptions of each topic. This was achieved using a custom-engineered prompt to maximize the clarity and informativeness of labels. The prompt used was as follows:

> *I have a topic that is described by the following keywords: [KEYWORDS]. In this topic, the following documents are a small but representative subset of all documents in the topic: [DOCUMENTS]. Based on the information above, please give a short label and an informative description of this topic in the following format: <label> ; <description> .*

This approach allowed us to produce interpretable labels while preserving the semantic structure of the topics.

**Appendix B: Details Regarding Human Annotation Data**

In this section, we briefly outline the procedure used to annotate the bodyworn camera recordings. More details on this process, including annotator training materials and full text of the annotation manuals, are available via Open Science Framework (anonymized for review) https://osf.io/py6md/overview?view_only=bcf30af302ff4c19aaf6b61300fc71ba.

The LAPD allowed our team access to nearly 2,000 bodyworn camera recordings associated with a stratified random sample of approximately 1,000 motor vehicle stops conducted by LAPD officers between 9/01/2021 and 8/31/2022. The sample is stratified by month and balanced to include 50% stops that involve a search and 50% non-search stops. Over 40 annotators were recruited and trained, including former law enforcement officers, individuals with a history of past arrest, and individuals of varying age, race, gender, neighborhood of residence, and profession. These annotators coded data on each bodyworn camera video using custom annotation software developed by our team, called CVAT-BWV (Hejabi et al. 2025).

Before reviewing LAPD bodyworn camera footage, annotators undergo a five-hour training program on the platform. The training walks through the full annotation manual (Trager et al. 2024) and the Bank of Labels, incorporates supervised, real-time annotation practice with publicly available bodyworn camera videos, and introduces resources and strategies for recognizing, mitigating, and coping with secondary trauma.

The annotation procedure proceeded in multiple phases. In Phase 0, annotators corrected errors in automated transcription and diarization. In Phase 1A, annotators identified and tagged police officers, civilians, and key objects in bodyworn camera footage. Phase 1B involved labeling the demographic characteristics (e.g., race, gender, and age) of the individuals identified in Phase 1A. Phase 2 focused on evaluating both objective and subjective aspects of officer–civilian interactions. Phase 2A captured emotions and other subjectively assessed elements, along with key actions and outcomes. Phase 2B measured the level of respect displayed by officers and civilians at 30-second intervals and recorded overall assessments of the stop.

Our annotation process produced 5,941 annotations across all three phases, covering 1,477 videos, 899 stops, and involving 44 annotators.[3]

---

[3] Our team's data use agreement with the LAPD gives us access to videos from a stratified random sample of 1000 traffic stops. However, some of the stops in the sample either turned out not to be motor vehicle stops or the videos associated with those stops did not capture enough information about the interaction to enable analysis. For

In this paper, we focus primarily on subjective items from Phase 2B and two composite measures constructed from Phase 2A annotation items. Table A1 provides substantive descriptions and summary statistics for all items from multiple annotated videos. Phase 2B items were measured on a five-point Likert scale. Phase 2A items were coded separately for each driver or passenger appearing in the video, with annotators able to respond "Yes," "No," "Not obvious from video," or "Uncertain." We treat responses other than "Yes" or "No" as missing. For each Phase 2A item, we then construct a binary indicator that equals 1 if any annotation for a given stop was coded "Yes," and 0 otherwise.[4]

The unit of analysis in this paper is at the annotation level, and our analyses are conducted separately by phase. For example, the variance decomposition for the *Respect: Officer* variable is conducted on annotations from phase 2b, but only those from videos with multiple annotations. The variance decomposition for *Any Person Searched*, on the other hand, is conducted on annotations from phase 2a, again using annotations from videos with multiple annotations, and only includes annotations with explicit answers. This variable captures searches of a person and excludes patdowns or vehicle searches. The analysis for *Any Contraband Found* is conducted the same way, but this question was only asked of annotators conditional on a search being reported. As such, *Any Contraband Found* has fewer observations.

***Table A1: Annotation Variable Description***

| **Variable** | **Description** | **Mean** | **SD** | **N** |
|---|---|---|---|---|
| *Danger: Driver* | How in danger or threatened do you think the Driver felt during this traffic stop? | 1.78 | 0.995 | 809 |
| *Danger: Officer* | How in danger or threatened do you think the Primary Officer felt during this traffic stop? | 1.38 | 0.737 | 811 |

example, sometimes the only video available is from an officer who did not interact with the driver and whose camera did not capture the main interaction.

[4] Note that the mean for searches is lower than the mean for contraband, as other forms of searches, such as pat-downs and vehicle searches, are not captured by this item, and the contraband items are asked conditional on a search being reported.

| *Respect: Driver* | Overall, how would you evaluate the Driver’s respectfulness during this stop | 4.01 | 0.785 | 809 |
|---|---|---|---|---|
| *Respect: Officer* | Overall, how would you evaluate the Primary Officer’s respectfulness during this stop | 4.00 | 0.846 | 811 |
| *Stop: Legitimacy* | Taken in its entirety, does this stop represent legitimate policing? | 4.06 | 1.05 | 810 |
| *Any Person Searched?* | Did the annotator report that *any*Driver or passenger had been searched? (Excluding patdowns) | 0.078 | 0.269 | 1186 |
| *Any Contraband Found?* | Conditional on any search, did the annotator report that *any* contraband was found? | 0.138 | 0.345 | 559 |

## Appendix References